\documentclass{article}
\pdfoutput=1

\usepackage{graphicx,latexsym,amssymb,amsmath,euscript}

\begin{document}
\newtheorem{theorem}{Theorem}
\newtheorem{definition}{Definition}
\newtheorem{lemma}{Lemma}
\newenvironment{proof}{{\bf Proof}}{\hfill $\Box$}

\title{Perfect simulation of spatial point processes\\ using dominated coupling from the past with\\ application to a multiscale area-interaction\\ point process}

\author{Graeme K. Ambler\\University of Bristol \and Bernard W. Silverman\\University of Oxford}
\date{2004}

\maketitle

\begin{abstract}
We consider perfect simulation algorithms for locally stable point processes 
based on dominated coupling from the past.  A version of the algorithm is 
developed which is feasible for processes which are neither purely attractive 
nor purely repulsive.  Such processes include multiscale area-interaction 
processes, which are capable of modelling point patterns whose clustering 
structure varies across scales.  We prove correctness of the algorithm and 
existence of these processes.  An application to the redwood seedlings data 
is discussed.
\end{abstract}


\section{Introduction}

One of the long standing problems in Markov chain Monte Carlo is that it is 
rarely possible to know when the Markov chain we are using for simulation has 
reached equilibrium.  For certain classes of problem, this problem was solved 
by the introduction of coupling from the past (CFTP) 
\cite{pro-wil:exa,pro-wil:unk-mar}.  More recently, methods based on 
CFTP have been developed for perfect simulation of spatial point process models 
(see for example \cite{ken:are,ken:boo,hag-lie-mol:exa-spa,ken-mol:per}).

Exact CFTP methods are therefore attractive, as one does not need to rigorously 
check convergence or worry about burn-in, or use complicated methods to find 
appropriate standard errors for Monte Carlo estimates based on correlated samples.  
Independent and identically distributed samples are now available, so estimation 
reduces to the simplest case.  Unfortunately, this simplicity comes at a price.  
These methods are notorious for taking a long time to return just one exact 
sample and are often difficult to code, leading many to give up and return to 
nonexact methods.

In response to these issues, in the first part of this paper we present a 
dominated CFTP algorithm for the simulation of locally stable point processes 
which potentially requires far fewer evaluations per iteration than the existing 
method in the literature \cite{ken-mol:per}.

It is often the case that advancements in statistical theory are inspired by 
applications, and this is no exception.  There are several classes of model for 
stochastic point processes, for example simple Poisson processes, cluster 
processes such as Cox processes, and processes defined as the stationary 
distribution of Markov point processes, such as Strauss processes 
\cite{str:a-mod} and area-interaction processes \cite{bad-lie:are-int}.  

All of the above mentioned point process models are capable of modelling either 
clustered or regular point patterns.  They are not, however, well suited to 
modelling point patterns whose clustering structure varies across scales, for 
example clusters of regularly spaced points or regularly spaced clusters of 
points.  In the second part 
of the paper we introduce a new multiscale area-interaction process which is 
capable of modelling either of these types of point pattern.  We then 
demonstrate how the algorithm developed in the first part of the 
paper may be used to generate samples from this process.

The structure of this papers is as follows.  In Section~\ref{simulation} 
we discuss perfect simulation, beginning with ordinary 
coupling from the past (CFTP) and moving on to dominated CFTP for spatial point 
processes.  We then introduce and justify our perfect simulation algorithm.  In 
Section \ref{aiprocess} we first review the standard area-interaction process.  We 
then introduce our multiscale process, describe how to use our new perfect 
simulation algorithm to simulate from it, and discuss a method for inferring 
the parameter values from data.  An application to the Redwood seedlings data is 
presented in Section \ref{appl}, and some areas for future work are discussed in 
Section~\ref{discuss}.

\section{Perfect simulation}\label{simulation}
\subsection{Coupling from the past}\label{cftp}

The principle behind CFTP is the following.  Suppose that it is desirable to 
sample from the stationary distribution of an ergodic Markov chain $\{Z_t\}$ 
on some (finite) state space $X$ with states $1,\ldots,n$.  It is clear that 
if it were possible to go back an infinite amount in time, start the chain 
running (in state $Z_{-\infty}$) and then return to the present, the chain 
would (with probability 1) be in its stationary distribution when one 
returned to the present (i.e. $Z_0\sim\pi$, where $\pi$ is the stationary 
distribution of the chain).

Suppose now that we were to set not one, but $n$ chains 
$\{Z^{(1)}_t\}, \ldots, \{Z^{(n)}_t\}$ running at a fixed time $-M$ in the 
past, where $Z^{(i)}_{-M}=i$ for each chain $\{Z^{(i)}_t\}$.  Now let all 
the chains be coupled so that if $Z^{(i)}_{s}=Z^{(j)}_{s}$ at any time $s$ 
then $Z^{(i)}_{t}=Z^{(j)}_{t} \hspace{1em} \forall t\geq s$.  Then if 
all the chains ended up in the same state $j$ at time zero 
(i.e. $Z^{(i)}_0=j \hspace{1em} \forall i\in X$), we would know that 
whichever state the chain passing from time minus infinity to zero was in 
at time $-M$, the chain would end up in state $j$ at time zero.  Thus $j$ 
must be a sample from the stationary distribution of the Markov chain in 
question.

When performing CFTP, a useful property of the coupling chosen is that it be
\emph{stochastically monotone} as in the following definition.

\begin{definition}\label{sto-mono}
Let $\{Z^{(i)}_t\}$ and $\{Z^{(j)}_t\}$ be two Markov chains obeying the 
same transition kernel.  Then a coupling of these Markov chains is 
stochastically monotone with respect to a partial ordering $\leq$ if 
whenever $Z^{(i)}_t\leq Z^{(j)}_t$, then $Z^{(i)}_{t+k}\leq Z^{(j)}_{t+k}$ 
for all positive $k$.
\end{definition}

Whenever the coupling used is stochastically monotone and there are maximal 
and minimal elements with respect to $\leq$ then we need only simulate 
chains which start in the top and bottom states, since chains starting in all 
other states are sandwiched by these two.  This is an important ingredient 
of the dominated coupling from the past algorithm introduced in the next 
section.

Although 
attempts have been made to generalise CFTP to continuous state 
spaces (notably \cite{mur-gre:con} and \cite{gre-mur:bay}, 
as well as \cite{ken-mol:per}, discussed in Section~\ref{domcftp}), 
there is still much work to be done before exact sampling becomes 
universally, or even generally applicable.  For example, there are no truly 
general methods for processes in high, or even moderate, dimensions.

\subsection{Dominated coupling from the past}
\label{domcftp}

Dominated coupling from the past was introduced as an extension of coupling 
from the past which allowed the simulation of the area-interaction process 
\cite{ken:are}, though it was soon extended to other types of point processes 
and more general spaces \cite{ken-mol:per}.  We give the formulation for 
locally stable point processes.

Suppose that we wish to obtain a sample of a spatial point process with density 
$f$ with respect to the unit rate Poisson process, whose Papangelou 
conditional intensity, $\lambda_f(u;X)$, is uniformly bounded above by some 
constant $\lambda$:
\[
	 \lambda_f(u;X)=\frac{f(X\cup\{u\})}{f(X)}\leq\lambda.
\]
The uniform bound on the Papangelou conditional intensity is required in order 
for the point process to be locally stable, so this is not imposing any 
additional constraint.

Then the algorithm given in \cite{ken-mol:per} is as follows.

\begin{enumerate}
\item Obtain a sample of the Poisson process with rate $\lambda$.
\item Evolve a Markov process $D(T)$ \emph{backwards} until some fixed time $-T$, 
using a birth-and-death process with death rate equal to 1 and birth rate equal 
to $\lambda$.  The configuration generated in step 1 is used as the initial state.
\item Mark all of the points in the process with U[0,1] marks.  We refer to the 
mark of point $x$ as $P(x)$.
\item \label{init} Recursively define upper and lower processes, $U$ and $L$ as 
follows.  The initial configurations at time $-T$ for the processes are 
\begin{eqnarray*}
	U_{-T}(-T) & = & \left\{x\; : x\in D(-T)\right\}\\
	L_{-T}(-T) & = & \left\{\boldsymbol{0}\right\}
\end{eqnarray*}
\item \label{evolve} Evolve the processes \emph{forwards} in time to $t=0$ in the 
following way.

Suppose that the processes have been generated up a given time, $u$, and suppose that 
the next birth or death to occur after that time happens at time $t_i$.  If 
a \textbf{birth} happens next then we accept the birth of 
the point $x$ in $U_{-T}$ or $L_{-T}$ if the point's mark, $P(x)$, is less than
\begin{eqnarray}
\nonumber	& \min\left\{\frac{\lambda_f(x;X)}{\lambda}: 
		L_{-T}(t_i)\subseteq X\subseteq U_{-T}(t_i)\right\} &\text{ or}\\
\label{birthrate} &	\max\left\{\frac{\lambda_f(x;X)}{\lambda}: 
		L_{-T}(t_i)\subseteq X\subseteq U_{-T}(t_i)\right\} &
\end{eqnarray}
respectively, where $x$ is the point to be born.

If, however, a \textbf{death} happens next then if the event is present 
in either of our processes we remove the dying event, setting
$U_{-T}(t_i) = U_{-T}(u)\setminus \{x\}$ and 
$L_{-T}(t_i) = L_{-T}(u)\setminus \{x\}$.
\item Define $U_{-T}(u+\varepsilon) = U_{-T}(u)$ and 
$L_{-T}(u+\varepsilon) = L_{-T}(u)$ for $u<u+\varepsilon<t_i$.
\item If $U_{-T}$ and $L_{-T}$ are identical at time zero 
(i.e. if $U_{-T}(0) = L_{-T}(0)$), then we have the required sample 
from the area-interaction process with rate parameter $\lambda$ and 
attraction parameter $\gamma$.  If not, go to step 2 and repeat, extending 
the underlying Poisson process back to $-(T+S)$ and generating 
additional $U[0,1]$ marks (keeping the ones already generated).
\end{enumerate}

This algorithm involves calculation of $\lambda(u;X)$ for each 
configuration that is both a subset of $U(T)$ and a superset of $L(T)$.  
Since calculation of $\lambda(u;X)$ is typically expensive, this calculation 
may be very costly.  The method proposed in Section~\ref{perfect_alg} uses 
an alternative version of step~\ref{evolve} which only requires us to 
calculate $\lambda(u;X)$ for upper and lower processes.

The more general form given in \cite{ken-mol:per} may be obtained from the 
above algorithm by replacing the evolving Poisson process $D(T)$ with a 
general dominating process on a partially ordered space $(\Omega,\preceq)$ 
with a unique minimal element $\boldsymbol{0}$.  The partial ordering in the 
above algorithm is that induced by the subset relation $\subseteq$.  
Step~\ref{evolve} is replaced by any step which preserves the sandwiching 
relations
\begin{eqnarray}
\label{sand1} &	L_{-T}(u)\preceq X_{-T}(u)\preceq U_{-T}(u)\preceq D(u) &
\text{and}\\
\label{sand2} &	L_{-T}(t)=U_{-T}(t)\;\text{ if }\;L_{-T}(s)=U_{-T}(s) &
\end{eqnarray}
for $s\leq t\leq0$, and the funnelling property
\begin{equation}
\label{funnel}	L_{-T}(u)\preceq L_{-(T+S)}(u)\preceq U_{-(T+S)}(u)\preceq U_{-T}(u)
\end{equation}
for all $u<0$ and $T,S>0$.  In equation (\ref{sand1}), $X_{-T}(u)$ is the 
Markov chain or process from whose stationary distribution we wish to sample.

\subsection{A new perfect simulation algorithm}\label{perfect_alg}

Suppose that we wish to sample from a locally stable point process with density
\begin{equation}\label{gen-proc}
	p(X) = \alpha\prod_{i=1}^m f_i(X),
\end{equation}
where $\alpha\in(0,\infty)$ and $f_i:\mathfrak{R}^f\to\mathbb{R}$ are 
positive valued functions which are monotonic with respect to the 
partial ordering $\preceq$ induced by the subset relation\footnote{That is, 
configurations $x$ and $y$ satisfy $x\preceq y$ if $x\subseteq y$.} and 
have uniformly bounded Papangelou conditional intensity:
\[
	\lambda_{f_i}(u;\mathbf{x})
		=\frac{f_i(\mathbf{x}\cup\{u\})}{f_i(\mathbf{x})}
		\leq K.
\]

Then clearly
\begin{equation}\label{gen-dom}
	\lambda_p(u;\mathbf{x}) \leq \lambda = 
		\prod_{i=1}^m\max_{X,\{x\}}\lambda_{f_i}(x;X)
\end{equation}
for all $u$ and $\mathbf{x}$, and $\lambda$ is finite.  Thus we may use 
the algorithm in Section~\ref{domcftp} to simulate from this process 
using a Poisson process with rate $\lambda$ as the dominating process.

However, as previously mentioned, calculation of $\lambda_p(u;\mathbf{x})$ 
is typically expensive, increasing at least linearly in $n(\mathbf{x})$.  
Thus to calculate the expressions in display~(\ref{birthrate}), we must in 
general perform $2^{n(U_{-T}(t_i))-n(L_{-T}(t_i))}$ of these 
calculations, making the algorithm non-polynomial.  In practice it is 
clearly not feasible to use this algorithm in all but the most trivial of 
cases, so we must look for some way to reduce the computational burden in 
step~\ref{evolve} of the algorithm.

This can be done by replacing step~\ref{evolve} with the following alternative.

\begin{enumerate}
\item[\ref{evolve}'] Evolve the processes \emph{forwards} in 
time to $t=0$ in the following way.

Suppose that the processes have been generated up a given time, $u$, and suppose 
that the next birth or death to occur after that time happens at time $t_i$.  If 
a \textbf{birth} happens next then we accept the birth of 
the point $x$ in $U_{-T}$ or $L_{-T}$ if the point's mark, $P(x)$, is less than
\begin{eqnarray}
\label{gen-up}	& \prod_{i=1}^m \left[\max\left\{ \lambda_{f_i}(u;U(T)) , 
			\lambda_{f_i}(u;L(T))\right\}
		\left/ \lambda\right.\right] 
								&\text{ or}\\
\label{gen-low} & \prod_{i=1}^m \left[\min\left\{ \lambda_{f_i}(u;U(T)) , 
			\lambda_{f_i}(u;L(T))\right\}
		\left/ \lambda\right.\right] &
\end{eqnarray}
respectively, where $x$ is the point to be born.

If, however, a \textbf{death} happens next then if the event is present 
in either of our processes we remove the dying event, setting
$U_{-T}(t_i) = U_{-T}(u)\setminus \{x\}$ and 
$L_{-T}(t_i) = L_{-T}(u)\setminus \{x\}$.
\end{enumerate}

\begin{lemma}\label{cftp_lemma}
Step~\ref{evolve}' obeys properties~(\ref{sand1}), (\ref{sand2}) and 
(\ref{funnel}) and is thus a valid dominated coupling from the past 
algorithm.
\end{lemma}
\begin{proof} Property~(\ref{sand1}) follows by noting 
that
\[
(\ref{gen-low})\leq\lambda_p(u;X)\leq(\ref{gen-up})\leq1.
\]
Property~(\ref{sand2}) is trivial.  Property~(\ref{funnel}) follows 
from the monotonicity of the $f_i$'s.\hfill$\Box$
\end{proof}

\begin{theorem}
Suppose that we wish to simulate from a locally stable point process whose 
density $p(X)$ with respect to the unit rate Poisson process is representable 
in form~(\ref{gen-proc}).  Then by replacing Step~\ref{evolve} by 
Step~\ref{evolve}' it is possible to bound the necessary number of 
calculations of $\lambda_p(u;X)$ per iteration in the dominated 
coupling from the past algorithm independently of $n(X)$.
\end{theorem}

\begin{proof} Step~\ref{evolve}' clearly involves only a 
constant number of calculations, so by
Lemma~\ref{cftp_lemma} above and Theorem~2.1 of \cite{ken-mol:per} the 
result holds.\hfill$\Box$
\end{proof}

In the case where it is possible to write $p(X)$ in form~(\ref{gen-proc}) 
with $m=1$, Step~\ref{evolve}' is identical to Step~\ref{evolve}.  This is 
the case for models which are either purely attractive or purely repulsive, 
such as the standard area-interaction process discussed in 
Section~\ref{std_aip}.  It is not the case for the multiscale process 
discussed in Section~\ref{multi}, or the model studied in \cite{amb-sil:wave}.  

The proof of 
Theorem~2.1 in \cite{ken-mol:per} does not require that the initial 
configuration of $L_{-T}$ be the minimal element $\mathbf{0}$, only that 
it be constructed in such a way as properties (\ref{sand1}), (\ref{sand2}) 
and (\ref{funnel}) are satisfied.  Thus we may refine our method further by 
modifying step~\ref{init} so that the initial configuration of $L_{-T}$ is 
given by
\begin{equation}\label{gen-thin}
	L_{-T}(-T) = \left\{x\in D(-T):P(x)\leq\prod_{i=1}^m\left[\min_{X,\{x\}}
		\lambda_{f_i}(x;X)\left/\lambda\right.\right]\right\},
\end{equation}
which clearly satisfies the necessary requirements.

\section{Area-interaction processes}\label{aiprocess}

\subsection{Standard area-interaction process}\label{std_aip}

In the standard case, the area-interaction process has density 
\begin{equation}\label{aip}
	p(X) = \alpha\lambda^{N(X)}\gamma^{-m(X\oplus G)}
\end{equation}
with respect to the unit rate Poisson process, where $\alpha$ is a 
normalising constant, $\lambda>0$ is the \emph{rate} parameter, $N(X)$ is 
the number of points in the configuration $X$, $\gamma>0$ is the 
\emph{clustering} parameter, $G$ is some compact set in 
$\mathbb{R}^d$ and $X\oplus G$ 
is \emph{Minkowski addition}:
\[
	X\oplus G = \{a\in\mathbb{R}^d\mathrel{:}
		a=x+z\text{, where }x\in X\text{ and }z\in G\}.
\]
Here $0<\gamma<1$ is the \emph{repulsive} case, while $\gamma>1$ is the 
\emph{attractive} case.  The case $\gamma=1$ reduces the a homogeneous 
Poisson process with rate $\lambda$.

\begin{figure}
	\begin{center}
		\resizebox{0.6\textwidth}{!}{\hspace{-6em}\includegraphics{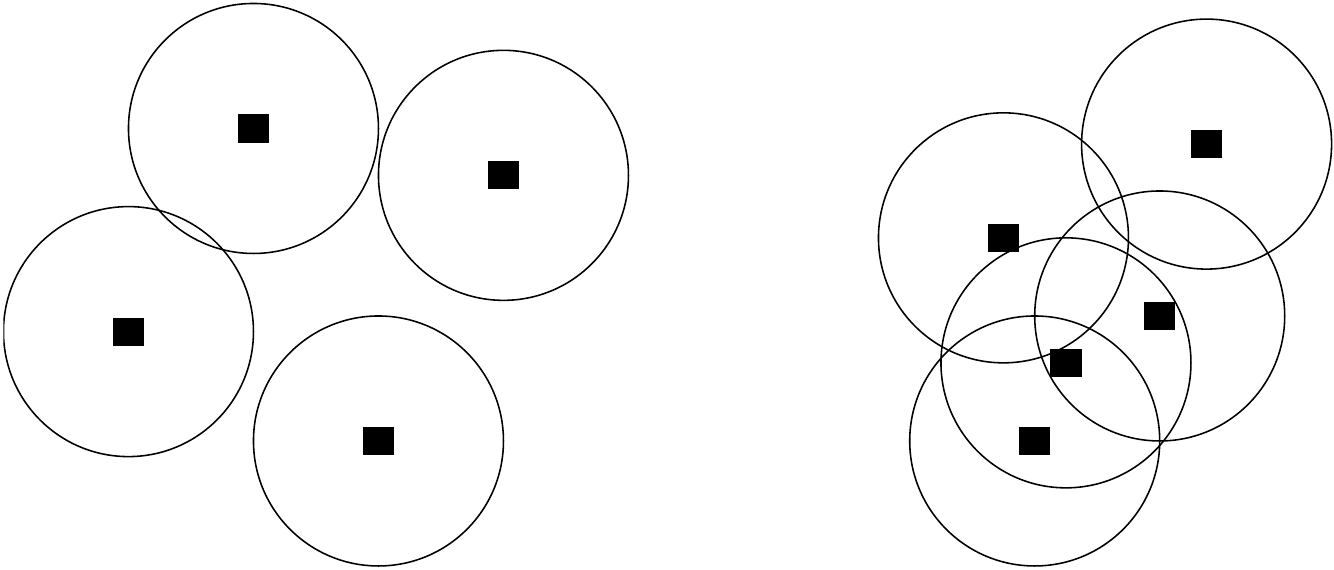}}
	\end{center}
		\caption[An example of some events together with circular ``grains'' G.]{An example of some events together with circular ``grains'' $G$.  The events in the above diagram would be the actual members of the process.  The circles around them are to show what the set $X\oplus G$ would look like.  If $\gamma$ were large, the point configuration on the right would be favoured, whereas if $\gamma$ were small, the configuration on the the left would be favoured.\label{fig:AIP}}
\vspace{1ex}
\end{figure}

Figure \ref{fig:AIP} gives an example of the construction when $G$ is a 
disc.

\subsection{A multiscale area-interaction process}\label{multi}

The area-interaction process is a flexible model which allows for a good 
range of models, from regular through total spatial randomness to 
clustered.  Unfortunately it does not allow for models whose behaviour 
changes at different resolutions, for example repulsion at small 
distances and attraction at large distances.  Some real-world examples of 
places where we see this sort of behaviour are the distribution of trees 
on a hillside, or the distribution of zebra in a patch of savannah.
A physical example of large scale attraction and small scale repulsion is 
the interaction between the strong nuclear force and the electro-magnetic 
force between two oppositely charged particles.  The physical 
laws governing this behaviour are different from those governing the 
behaviour of the area-interaction class of models, though they may be 
sufficiently similar so as to provide a useful approximation.

We propose the following model to capture these types of behaviour.

\begin{definition}
The \emph{multiscale area-interaction process} has 
density
\begin{equation}\label{att-rep}
	p(X) = \alpha\lambda^{N(X)}\gamma_1^{-m(X\oplus G_1)}
					\gamma_2^{-m(X\oplus G_2)},
\end{equation}
where $\alpha$, $\lambda$ and $N(X)$, are as in equation \eqref{aip}; 
$\gamma_1\in[1,\infty)$ and $\gamma_2\in(0,1]$; and
$G_1$ and $G_2$ are balls of radius $r_1$ and $r_2$ respectively.
\end{definition}

The process is clearly Markov of range $\max\{r_1,r_2\}$.  If 
$G_1\supset G_2$, we will have small scale repulsion and large scale 
attraction.  If $G_1\subset G_2$, we will have small scale attraction and 
large scale repulsion.

\begin{theorem}\label{thm:att-rep}
The density \eqref{att-rep} is both measurable and integrable.
\end{theorem}

This is a straightforward extension of the proof of Baddeley and van 
Lieshout \cite{bad-lie:are-int} for the standard area-interaction process.  
For completeness, it is given in Appendix~\ref{proof}.

\subsection{Perfect simulation of the multiscale process}\label{perfectmulti}

Perfect simulation of the multiscale process (\ref{att-rep}) is possible 
using the method introduced in Section~\ref{perfect_alg}.  Since 
(\ref{att-rep}) is already written as a product of three monotonic 
functions with uniformly bounded Papangelou conditional intensities, we need 
only substitute into equations (\ref{gen-dom}--\ref{gen-thin}) as follows.

Substituting 
into equation (\ref{gen-dom}), we find that the rate of a suitable 
dominating process is
\[
	\lambda\gamma_2^{-m(G_2)}.
\]
The initial configurations of the upper and lower process $U$ and $L$ are 
then found by simulating this process, thinning with a probability of 
\[
	\gamma_1^{-m(G_1)}\gamma_2^{m(G_2)}
\]
for $L$.

As $U$ and $L$ evolve towards time $0$, we accept points $x$ in $U$ with 
probability
\begin{equation}\label{multiUaccept}
		\gamma_1^{-m((x\oplus G_1)\setminus U_{-T}(u)\oplus G_1)}
		\gamma_2^{m(G_2)-m((x\oplus G_2)\setminus L_{-T}(u)\oplus G_2)}
\end{equation}
and accept events in $L$ whenever 
\begin{equation}\label{multiLaccept}
		P(x)\leq\gamma_1^{-m((x\oplus G_1)\setminus L_{-T}(u)\oplus G_1)}
		\gamma_2^{m(G_2)-m((x\oplus G_2)\setminus U_{-T}(u)\oplus G_2)}.
\end{equation}

\begin{figure}[t]
	\begin{center}
		\resizebox{0.9\textwidth}{!}{\includegraphics{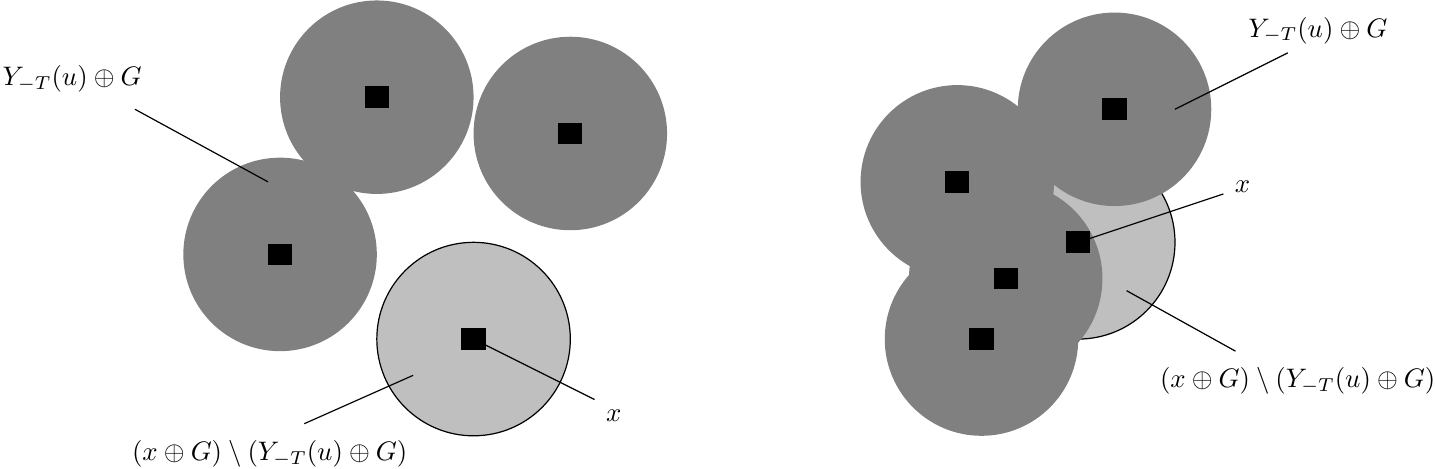}}
	\end{center}
		\caption[Another look at Figure \ref{fig:AIP}.]{Another look at Figure \ref{fig:AIP} with some shading added to show the process of simulation.  Dark shading shows $Y_{-T}(u)\oplus G$ where $Y_{-T}(u)$ is the state of either $U$ or $L$ immediately before we add the new event and $G$ could be either $G_1$ or $G_2$.  Light shading shows the amount added if we accept the new event.  In the configuration on the left, $x\oplus G=(x\oplus G)\setminus(Y_{-T}(u)\oplus G)$, so that the attractive term in (\ref{multiUaccept}) or (\ref{multiLaccept}) will be very small, whereas the repulsive term will be large.  In the configuration on the right we are adding very little area to $(Y_{-T}(u)\oplus G)$ by adding the event, so the attractive term will be larger and the repulsive term will be smaller.\label{fig:AIP2}}
\vspace{1ex}
\end{figure}

Figure \ref{fig:AIP2} 
gives examples of the construction $(x\oplus G)\setminus Y_{-T}(u)\oplus G$.

\subsection{Parametric inference}\label{arParEst}

We use maximum pseudo-likelihood 
\cite{bes:spa-int,bes:sta-ana,bes:met-sta,jen-mol:pse-exp} to 
estimate the parameters $\lambda$, $\gamma_1$ and $\gamma_2$.

As we saw in Section~\ref{perfectmulti}, the Papangelou conditional intensity 
of our process is
\[
	\lambda(u;X) = \lambda
		\gamma_1^{-m((u\oplus G_1)\setminus X\oplus G_1)}
		\gamma_2^{-m((u\oplus G_2)\setminus X\oplus G_2)}.
\]
Thus the pseudo-likelihood equations for this model are
\begin{align}\label{PseudoLambda}
	\sum_{x_i\in A} \frac1\lambda \; = \;& 
		\int_A \gamma_1^{-m((u\oplus G_1)\setminus X\oplus G_1)}
			\gamma_2^{-m((u\oplus G_2)\setminus X\oplus G_2)}du,\\
\nonumber
	\sum_{x_i\in A} \frac{m((x_i\oplus G_1)\setminus X\oplus G_1)}
							{\gamma_1} \; = \;& 
		\int_A m((u\oplus G_1)\setminus X\oplus G_1) \times\\[-2ex]
\label{PseudoGamma1}
		&  \hspace{1.6em}\lambda
			\gamma_1^{-m((u\oplus G_1)\setminus X\oplus G_1)} 
			\gamma_2^{-m((u\oplus G_2)\setminus X\oplus G_2)}du \\
\intertext{and}
\nonumber
	\sum_{x_i\in A} \frac{m((x_i\oplus G_2)\setminus X\oplus G_2)}
							{\gamma_2} \; = \;& 
		\int_A m((u\oplus G_2)\setminus X\oplus G_2) \times\\[-2ex]
\label{PseudoGamma2}
		&  \hspace{1.6em}\lambda
			\gamma_1^{-m((u\oplus G_1)\setminus X\oplus G_1)}
			\gamma_2^{-m((u\oplus G_2)\setminus X\oplus G_2)}du,
\end{align}
where we recall that $A$ is an arbitrary subset of the window in which we 
observe the point process.
Clearly the main difficulty is in estimating the integrals on the right 
hand side of equations \eqref{PseudoLambda} to \eqref{PseudoGamma2}.  
This problem may be tackled directly \cite{bad-tur:pra-max} by noting that 
the integral in the log pseudo-likelihood
\[
	\log\text{PL}(\theta;X) = \sum_{x_i\in A}\log\lambda(x_i;X) 
		- \int_A\lambda(u;X)du
\]
can be approximated by 
\[
	\int_A\lambda(u;X)du \simeq \sum_{j=1}^m\lambda(u_j;X)w_j,
\]
where $u_j$ are points in $A$ and $w_j$ are quadrature weights.  Using 
and extending an observation made by \cite{ber-tur:app-poi}, 
\cite{bad-tur:pra-max} note 
that if the set $\{u_j:j=1,\ldots,m\}$ contains all the events 
$\{x_i:i=1,\ldots,n(X)\}$, then the log pseudo-likelihood may be 
approximated by
\begin{equation}\label{logPL}
	\log\text{PL}(\theta;X) \simeq 
		\sum_{j=1}^m(y_j\log\lambda_j-\lambda_j)w_j,
\end{equation}
where $\lambda_j=\lambda(X,u_j)$, $y_j=z_j/w_j$ and 
\[
	z_j = \left\{\begin{array}{ccl}
			1 & \text{if} & u_j\in\{x_i:i=1,\ldots,n(X)\}\\
			0 & \text{if} & u_j\not\in\{x_i:i=1,\ldots,n(X)\}.
		     \end{array}\right.
\]
For a fixed point pattern $X$ the right hand side of \eqref{logPL} is 
equivalent to the log likelihood of independent Poisson variables 
$Y_k\sim\textrm{Poisson}(\lambda_k)$ taken with weights $w_k$, so \eqref{logPL} 
can therefore be maximised using standard software for fitting Generalised 
Linear Models, such as that in \textsf{R}.

In order to put the estimation procedure above into practice, we must 
have values for $r_1$ and $r_2$, the radii of $G_1$ and $G_2$ respectively.  
Following the lead of \cite{bad-tur:pra-max}, we suggest fitting the model 
for a variety of 
values of these ``nuisance parameters'' which do not fit into the 
exponential family model, and choosing the values which maximise the 
pseudo-likelihood.  It may be wise to plot estimates of some standard 
functions such as $K$ and $G$ in order to narrow the search somewhat.

\section{Redwood seedlings data}\label{appl}

\begin{figure}
 \begin{center}
  \resizebox{.95\textwidth}{!}{\hspace{-5em}\resizebox{.56\textwidth}{!}{\includegraphics{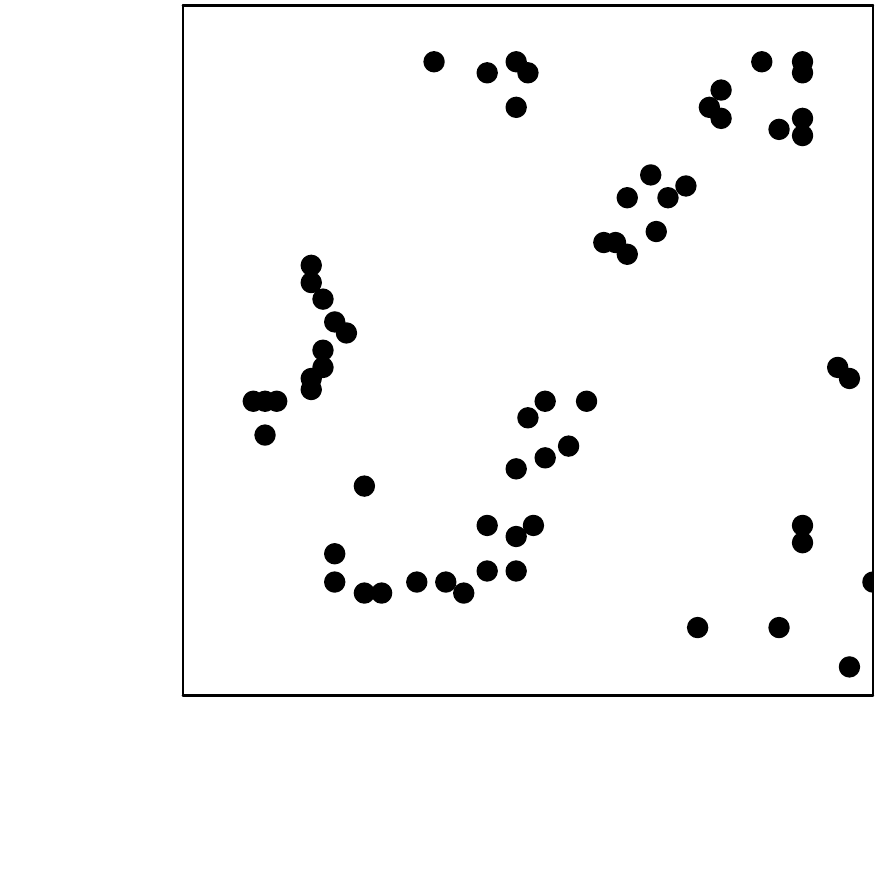}}\hspace{2em}\resizebox{.56\textwidth}{!}{\includegraphics{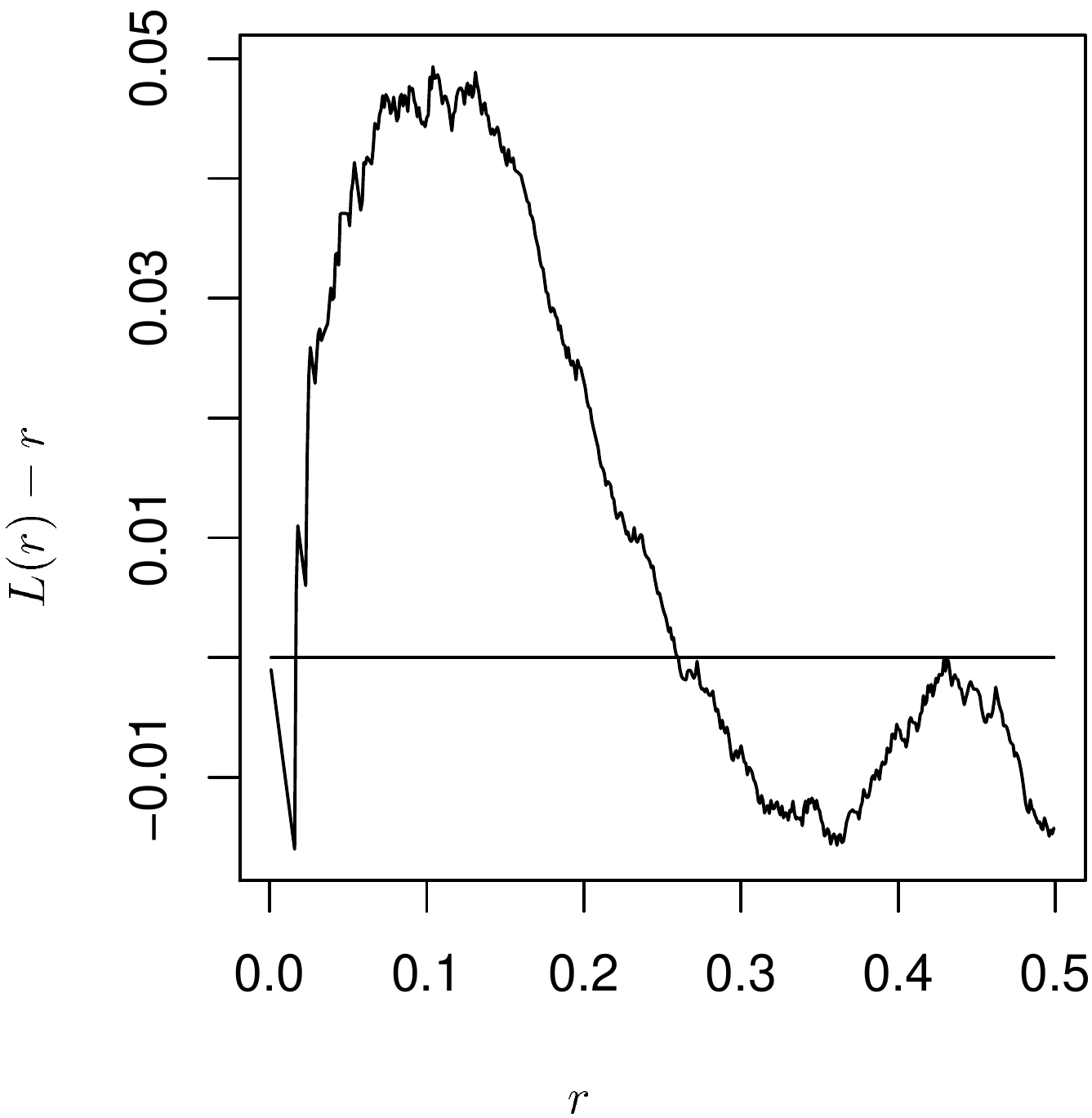}}}
 \end{center}
 \caption[Redwood seedlings data]{Redwood seedlings data.  Left: The data, selected by \cite{rip:mod-spa} from a larger data set analysed by \cite{str:a-mod}.  Right: Plot of the $L$ function for the redwood seedlings.  There seems to be interaction at 3 different scales: (very) small scale repulsion followed by attraction at a moderate scale and then repulsion at larger scales.}
 \label{fig:red-points}
\vspace{1ex}
\end{figure}

We take a brief look at a data set which has been much analysed in the 
literature, the Redwood seedlings data first considered by 
\cite{str:a-mod}.  We examine a subset of the original data chosen by 
\cite{rip:mod-spa} and later analysed by \cite{dig:par-est} among 
others.  The data are plotted in Figure \ref{fig:red-points}.  We wish to 
model this data using the multiscale model we have introduced.  
From an inspection of the estimated $K$-function (right pane in Figure 
\ref{fig:red-points}) of the data using Ripley's edge correction scheme 
\cite{rip:mod-spa} we estimate values of $R_1$ and $R_2$ as $0.07$ and 
$0.013$ respectively, giving repulsion at small scales and attraction at 
moderate scales.  It also seems that there is some repulsion at slightly 
larger scales, so it may be possible to use $R_2=0.2$ and to model the 
large scale interaction rather than the small scale interaction as we 
have chosen.

\begin{figure}
 \begin{center}
  \resizebox{.95\textwidth}{!}{\includegraphics{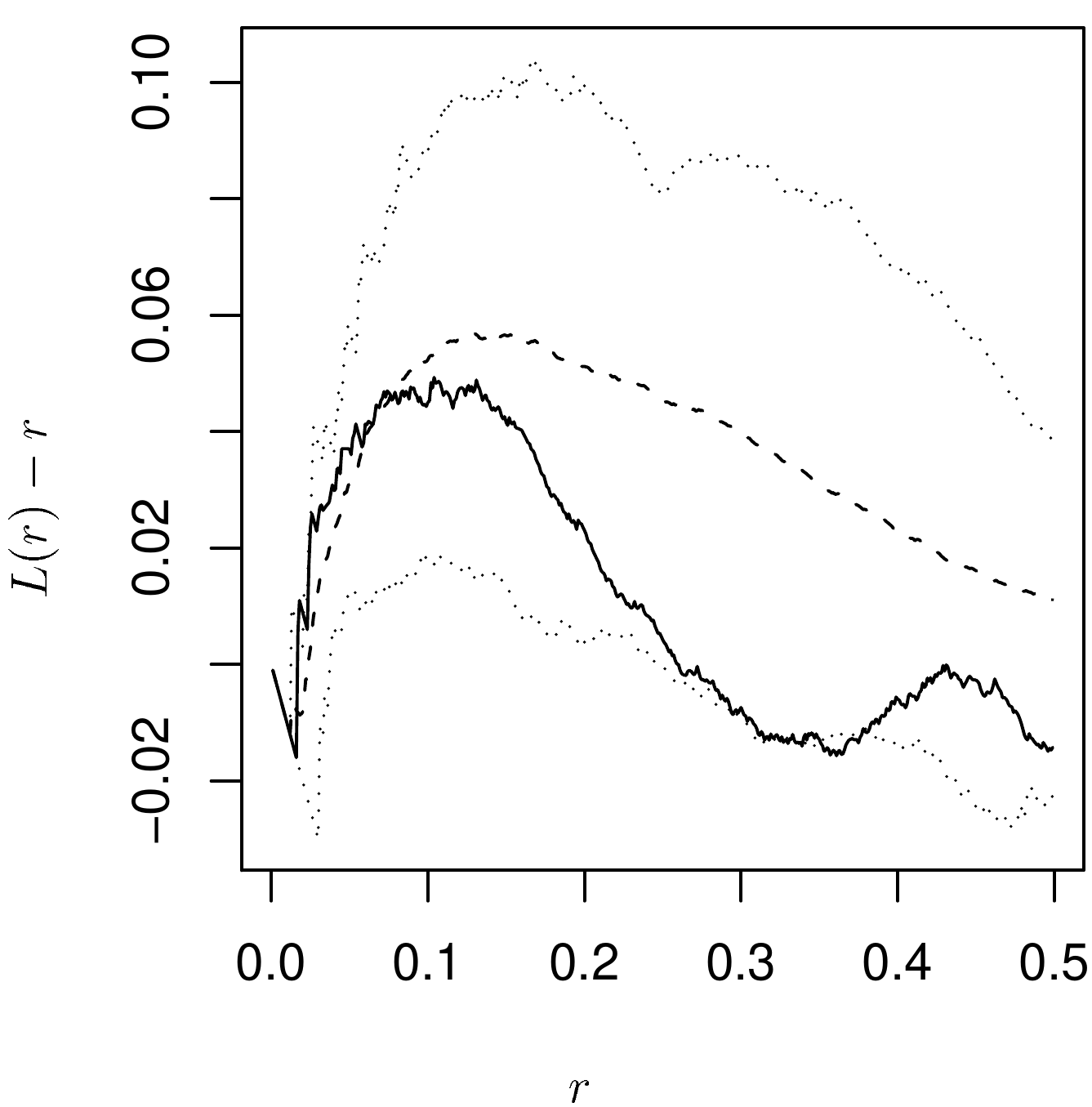}\hspace{2em}\includegraphics{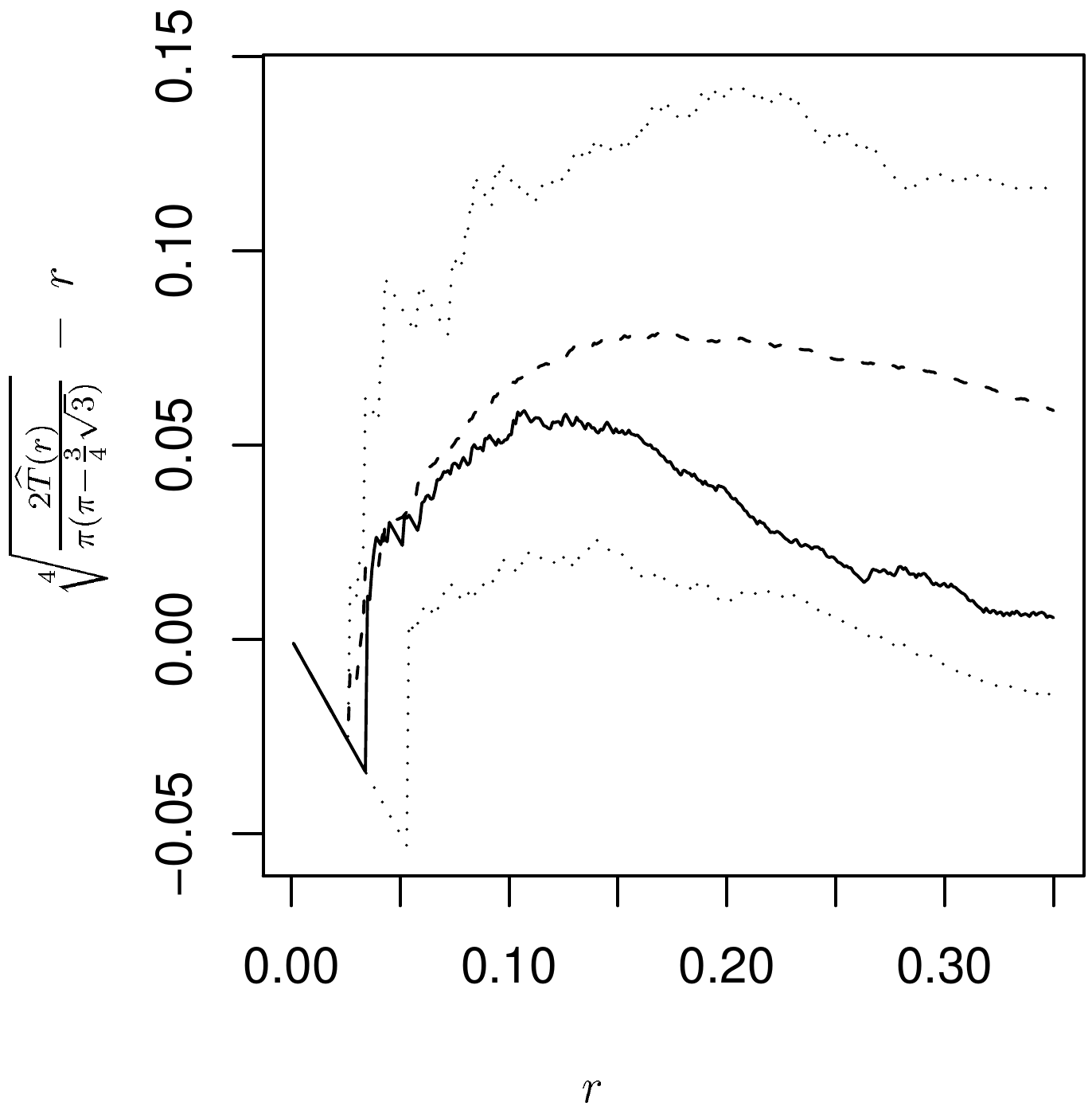}}
 \end{center}
 \caption[$L$ and $T$ function plots of the redwood seedlings data]{$L$ and $T$ function plots of the redwood seedlings data.  Left: $L$-function plots of the data together with simulations of the multiscale model with parameters $R_1=0.07$, $R_2=0.013$, $\lambda=0.118$, $\gamma_1=2000$ and $\gamma_2=10^{-200}$.  Dotted lines give an envelope of 19 simulations of the model, the solid line is the redwood seedlings data and the dashed line is the average of the 19 simulations.  Right: The same for the $T$ function.}
 \label{fig:simredk+t}
\vspace{1ex}
\end{figure}

Fitting the remaining parameters by eye again, we chose values 
$\lambda=0.118$, $\gamma_1=2000$ and $\gamma_2=10^{-200}$.  The remarkably 
small value of $\gamma_2$ was necessary because the value of $R_2$ was 
also very small.  It is clear from these numbers that it would be more 
natural to define $\gamma_1$ and $\gamma_2$ on a logarithmic scale.  
Figure \ref{fig:simredk+t} shows $K$ and $T$ function plots for 19 
simulations from this model, providing approximate 95\% Monte-Carlo 
confidence envelopes for the values of the functions.  It can be seen that 
on the basis of these functions, the model appears to fit the data 
reasonably well.

The plots show several things:  Firstly that the model fits reasonably well, 
but that it is possible that we chose a value of $R_1$ which was slightly too 
large.  Perhaps $R_1=0.06$ would have been better.  Secondly, it seems 
that the large scale repulsion may be an important factor which should not be 
ignored.  Thirdly, in this case we have gained little new information by 
plotting the $T$ function --- the third order behaviour of the data seems 
to be similar in nature to the second order structure.

\section{Discussion and future work}\label{discuss}

We have developed a new method for perfect simulation of locally stable point 
processes.  The main advantage of our method is that it allows acceptance 
probabilities to be computed in $O(n)$ instead of $O(2^{n})$ steps for models 
which are neither purely attractive nor purely repulsive.  Because of the 
exponential dependence on $n$, the algorithm of \cite{ken-mol:per} is not 
feasible in these situations.

We have also developed a multiscale area-interaction process 
which incorporates both repulsion and attraction and given a method of 
simulating this process exactly.  In addition, we have described a method of 
parametric inference for model fitting and given a small application to the 
redwood seedlings data \cite{str:a-mod}.

The practical work of the paper has focused on two-scale models but it is 
clear that in practice it is possible to extend the work to multiscale 
models in the general sense.  For example, the sample $L$-function of the 
redwood seedlings might, if the sample size were larger, indicate the 
appropriateness of a three scale model
\begin{equation}\label{threescale}
	p(X) = \alpha\lambda^{N(X)}\gamma_1^{-m(X\oplus G_1)}
					\gamma_2^{-m(X\oplus G_2)}
					\gamma_3^{-m(X\oplus G_3)}.
\end{equation}
The proof given in Appendix~\ref{proof} can easily be 
extended to show the existence of this process, and (\ref{threescale}) 
is also amenable to perfect simulation using the method of 
Section~\ref{perfect_alg}.  Because of the small size of the redwood 
seedlings data set a model of this complexity is not warranted, but 
the fitting of such models, and even higher order multiscale models in 
appropriate circumstances, would be an interesting topic for future 
research.

\subsection*{Acknowledgement}
The first author would like to thank Guy Nason and Paul Northrop 
for helpful discussions.

\bibliographystyle{apt}

\bibliography{exact,space,books}

\appendix
\section{Proof of Theorem~\ref{thm:att-rep}}\label{proof}

We prove a slightly more general result.

\begin{definition}
The generalised multiscale area-interaction process has 
density
\begin{equation}\label{gatt-rep}
	p(X) = \alpha\lambda^{N(X)}\gamma_1^{-\nu_1(U_1(X))}
					\gamma_2^{-\nu_2(U_2(X))},
\end{equation}
where $\alpha$, $\lambda$ and $N(X)$, are as in equation \eqref{aip}; 
$\gamma_1\in[1,\infty)$ and $\gamma_2\in(0,1]$; $\nu_1$ and $\nu_2$ are 
Borel regular measures; $Z_1$ and $Z_2$ are myopically continuous 
functions and $U_j=\bigcup_{x_i\in X}Z_j(x_i)$.
\end{definition}

\begin{theorem}\label{thm:gatt-rep}
The density \eqref{gatt-rep} is both measurable and integrable.
\end{theorem}

\begin{proof} If $\gamma_1=1$ then \eqref{gatt-rep} is simply the 
repulsive case of the 
area-interaction process and the result holds.  If 
$\gamma_2=1$ then \eqref{gatt-rep} is simply the attractive case of the 
area-interaction process and again the result holds.
We now consider the case $\gamma_1>1$ and $\gamma_2<1$.

Let $t>0$ and consider $V=\{X\in\mathfrak{R}^f:\nu_1(U_1(X))<t\}$.  We will 
show that $V$ is open in the weak topology with respect to the function 
$U_1$, and thus that $\nu_1(U_1(X))$ is weakly upper semicontinuous.  Since we 
know that upper or lower semicontinuous functions are measurable it is 
then a short road to showing that (\ref{gatt-rep}) is measurable.

Pick $X\in V$.  Then since $\nu_1$ is regular there is an open set 
$G\subset\chi$ containing $U_1(X)$ such that $\nu_1(G)<t$ as well.  Now clearly 
$X$ has no events in the set $H=\{x\in\chi:Z_1(x)\cap G^c\ne\phi\}$, and 
$Z_1(H)\subset J=\{K\in\EuScript{K}:K\cap G^c\ne\phi\}$, which is a closed set 
in the myopic topology.  Clearly $Z_1^{-1}(J)=H$, and since $Z_1$ is a myopically 
continuous function this shows that $H$ is also closed.  It is now easy to see 
that $W=\{Y\in\mathfrak{R}^f:N(Y_H)=0\}$ (where $Y_H$ is the restriction of 
the configuration $Y$ to the set $H$) is open in the weak topology with 
respect to the function $U_1$, and since $V$ is the union of a collection of 
sets of the form $W$ then $V$ is also open.

This shows that $X\to\nu_1(U_1(X))$ is weakly upper semicontinuous.  Thus the 
map $X\to\exp(-\nu_1(U_1(X))\log\gamma_1)$ is weakly lower semicontinuous.  
Thus $X\to\exp(-\nu_1(U_1(X))\log\gamma_1)$ is measurable.

By a similar argument, the map $X\to\exp(-\nu_2(U_2(X))\log\gamma_2)$ is 
weakly upper semicontinuous, and thus $X\to\exp(-\nu_2(U_2(X))\log\gamma_2)$ is 
measurable.  Since 
$\lambda^{N(X)}$ is clearly measurable this means that (\ref{gatt-rep}) is 
measurable.

To see that \eqref{gatt-rep} is integrable note that 
\begin{eqnarray*}
	 &0 \leq \nu_1(U_1(X)) \leq \nu(\chi) < \infty & \text{ and}\\
	 &0 \leq \nu_2(U_2(X)) \leq \nu(\chi) < \infty. & 
\end{eqnarray*}
Now the function $f(X)=\lambda^{N(X)}$ is integrable, since this is simply 
the Radon-Nikod\'ym derivative of the Poisson process with rate $\lambda$ 
with respect to the unit rate Poisson process.  Hence \eqref{gatt-rep} is 
dominated by an integrable function and is therefore integrable.  In 
fact this shows the stronger result that the generalised area-interaction 
process measure is uniformly absolutely continuous with respect to the 
$\lambda$-rate Poisson process measure and so its Radon-Nikod\'ym derivative 
is uniformly bounded.\hfill$\Box$
\end{proof}

\end{document}